\documentclass[twocolumn,showpacs,superscriptaddress,prl]{revtex4} 
 
\usepackage{dcolumn} 
\usepackage{amssymb} 
\usepackage{amsmath} 
\usepackage{amsbsy} 
\usepackage{amsfonts} 
\newtheorem{theo}{Theorem} 
\newtheorem{lm}{Lemma} 
\newtheorem{co}{Corollary} 
\newtheorem{pr}{Proposition}

\newcommand{\ket}[1]{\left \vert #1 \right \rangle}

\newcommand{\vph}{\varphi} 
 
\newcommand{\qed}{\mbox{\rule{1.6mm}{2.3mm}}}

\begin{document} 
 
\title[Short Title]{Supercatalysis} 
 
\author{Somshubhro Bandyopadhyay} 
\email{som@ee.ucla.edu} 
\affiliation{Electrical Engineering Department,  UCLA, Los Angeles, CA 90095} 
\author{Vwani Roychowdhury} 
\email{vwani@ee.ucla.edu} 
\affiliation{Electrical Engineering Department,  UCLA, Los Angeles, CA 90095}

\date{\today}
 
\begin{abstract} 
We show that entanglement-assisted transformations of bipartite entangled  
states can be more efficient than catalysis [D. Jonathan and M. B. Plenio, Phys. Rev. Lett. \textbf{83}, 3566
(1999)], i.e., given two
incomparable bipartite   states not only can the transformation be enabled by 
performing collective operations with an auxiliary entangled state, but the
entanglement of the auxiliary  state itself can be enhanced. We refer to this
phenomenon as {\em supercatalysis}.   We  provide results on the properties of
supercatalysis and its relationship with catalysis. In particular,  we  
obtain a useful necessary and sufficient condition for catalysis,  and provide
several sufficient conditions for supercatalysis and study the   extent to
which entanglement of the auxiliary state can be enhanced via supercatalysis. 
\end{abstract}    
\maketitle 
One of the primary goals of quantum information theory \cite{qinfo} is efficient 
manipulation of quantum entanglement shared among spatially separated parties, 
each of whom possesses only a subsystem of the entire entangled state \cite{ent}. 
Such distributed entanglement, as a resource, 
is a critical component of  novel quantum information protocols, such as quantum teleportation 
\cite{tel}, super dense coding \cite{sdc} and of  distributed computing algorithms 
\cite{compu}. Since the underlying entangled state is spatially distributed, any entanglement 
manipulation is necessarily constrained to be carried out with  local operations and classical  
communication among the parties (LOCC). The properties and classifications 
of both  deterministic and probabilistic/conclusive LOCC transformations have 
been pursued vigorously in the recent past \cite{lp}-\cite{daftuar}.  
 
A surprising feature that sets apart entanglement from usual 
physical resources is its capacity to enable, without being consumed,  transformations 
that are impossible under {\em deterministic} LOCC \cite{jp}. This property is very similar to  
that of catalysts in chemical reactions and is aptly termed as {\em entanglement catalysis}.  
It has also been shown  that the probability 
of a conclusive conversion can  be enhanced in the presence of a catalyst, when 
a deterministic conversion is not possible \cite{jp}. Another instance where 
entanglement is useful  in a sense similar to catalysis (i.e., not being consumed), 
is {\em partial recovery} of entanglement. In this case, the entanglement lost in an LOCC  
manipulation is partially recovered 
using an auxiliary entanglement and performing collective operations \cite{mori,som}. 
 
We show for the first time that the above two features of entanglement  
can be exploited simultaneously and that entanglement assisted LOCC (ELOCC)  
transformations can be more efficient than catalysis \cite{jp}.  
In particular, given two incomparable states (i.e., states that are 
not LOCC transformable with certainty), not only can the transformation be enabled by 
performing collective operations with an auxiliary entangled state, but the entanglement of the auxiliary 
state itself can be enhanced. Such simultaneous enabling of deterministic-LOCC 
 impossible transformations, and   reduction of  the overall loss in entanglement 
is not possible under catalysis \cite{jp}. We refer to this phenomenon as {\em supercatalysis}. 
In this letter, we study the properties of supercatalysis and its relationship with catalysis,  
obtain a useful and succinct necessary and sufficient condition for catalysis, 
and sufficient conditions for supercatalysis. 
 
All the transformations that we consider in this letter are deterministic, i.e., 
occur with probability one, and are  in the finite copy regime. We represent an  
$n\times n$ bipartite  pure entangled state, $\ket{\psi}$, as   
$\ket{\psi}= \displaystyle \sum ^{n}_{i=1}\sqrt{\alpha
_{i}}\ket{i}\ket{i}$,   where $\alpha_1 \geq \alpha_2 \geq \cdots \geq
\alpha_{n}$, are the  Schmidt coefficients or eigenvalues of the reduced
density matrices. Also, let   $\lambda_{\psi}$ denote the vector of the
ordered eigenvalues.   Then, it follows from Nielsen's result \cite{nielsen}
that for any two given $n \times n$ states  $\ket{\psi} =\sum
^{n}_{i=1}\sqrt{\alpha _{i}}\ket{i}\ket{i}$  and $\ket{\vph}=\sum
^{n}_{j=1}\sqrt{\beta _{j}}\ket{j}\ket{j},$   $\ket{\psi}  \longrightarrow
\ket{\vph}$ with probability one under LOCC, if and  only if ,
$\lambda_{\psi}$ is   {\em majorized}\/ by $\lambda_{\vph}$, 
(denoted as $\lambda_{\psi} \prec \lambda_{\vph}$); i.e.,  
\begin{equation} 
\sum ^{m}_{i=1}\alpha_i \leq \sum ^{m}_{i=1}\beta_i, 
  \qquad \mbox{for every $m=1,\ldots,n-1$\ .} 
\label{eq:majorization} 
\end{equation} 
Note that the above inequality is satisfied trivially when $m=n$, since  
both sides equal 1. In the rest of the paper, for the sake of convenience, instead 
of representing a bipartite state $\ket{\psi}= \sum ^{n}_{i=1}\sqrt{\alpha
_{i}}\ket{i}\ket{i}$, we shall represent  it simply by the vector of its
eigenvalues: $\ket{\psi}=\left( \alpha_1, \alpha_2, \ldots, \alpha_{n} 
\right)$.

Consider the following pair of $4 \times 4$ {\em bipartite incomparable} states, 
\begin{eqnarray} 
\left| \psi \right\rangle  & = & \left(0.4, 0.36,  0.14, 0.1\right) \ ,\label{eq:1} \\ 
\left| \phi \right\rangle  & =  & \left(0.5, 0.25,  0.25, 0.0\right) \ ,\label{eq:2} 
\end{eqnarray} 
for which an auxiliary entangled state \( \left| \chi \right\rangle =\left(0.65, 0.35 \right)  \) 
is a catalyst, i.e., the transformation \( \left| \psi \right\rangle \otimes \left| \chi \right\rangle \rightarrow \left| \phi 
\right\rangle \otimes \left| \chi \right\rangle  \) 
can be realized deterministically under LOCC.  
An example of supercatalysis 
lies in showing the existence of a state, say $\ket{\omega}= \sum
^{k}_{i=1}\sqrt{\gamma _{i}}\ket{i}\ket{i}$,  such that \( \left| \psi
\right\rangle \otimes \left| \chi \right\rangle \rightarrow \left| \phi
\right\rangle \otimes \left| \omega  \right\rangle  \)  by LOCC with
probability one, where \( E\left( \omega \right) >E\left( \chi \right)  \), 
\( E \) being the entropy of entanglement (e.g.,   $E\left( \ket{\omega}
\right)=  -\sum_{i=1}^{k} \gamma_i \ln (\gamma_i)$). Let \( \left| \omega
\right\rangle =\left(0.55,   0.45\right) \). 
Note that \( E\left( \omega \right) >E\left( \chi \right)  \). The corresponding 
eigenvalue vectors $\lambda _{\psi \otimes \chi }, \lambda _{\phi \otimes
\omega }$ are given by  
\begin{eqnarray}  \left (0.26,0.234,0.14,0.126,0.091,0.065,0.049,0.035\right),
\label{3} \\ \left (0.275,0.225,0.1375,0.1375,0.1125,0.1125,0,0\right) \; \;
\; \;\;  \label{4}  \end{eqnarray}   respectively. It can be easily verified
that \( \lambda _{\psi \otimes \chi }\prec \lambda _{\phi \otimes \omega },
\)  and hence,  the transformation \( \left| \psi \right\rangle \otimes \left|
\chi  \right\rangle \rightarrow \left| \phi \right\rangle \otimes \left| \omega
\right\rangle  \)  is possible under LOCC with certainty.    As the final
state, $\ket{\omega}$,  of the initial auxiliary state, ($\ket{\psi}$),   is
more entangled than its initial one, supercatalysis  is clearly more efficient
than catalysis. An equivalent interpretation of  the underlying phenomenon is
that supercatalysis, in addition to enabling the transformation,  reduces the
overall loss in entanglement. In catalysis the net entanglement lost  is just
the difference between entanglement of the parent states. Supercatalysis 
reduces this loss by an amount \( \delta =E\left( \omega \right) -E\left( \chi
\right)  \).    One can think of several innovative uses of supercatalysis,
and a particular scenario,  where resources are limited and constrained is
outlined next.   For instance,  consider a scenario where we are given two
copies of the source  state, say \( \left| \psi \right\rangle =\left( 0.4,
0.4, 0.1, 0.1\right)  \)  and we wish to obtain the target states, \( \left|
\phi _{1}\right\rangle =\left( 0.5, 0.25, 0.25, 0\right)  \)  and \( \left|
\phi _{2}\right\rangle =\left( 0.48, 0.27, 0.25, 0 \right)  \)  respectively.
One can easily verify that all the following  pairs are  incomparable: \(
\left\{ \left| \psi \right\rangle ,\left| \phi _{1}\right\rangle \right\} 
\),   \( \left\{ \left| \psi \right\rangle ,\left| \phi _{2}\right\rangle
\right\}  \), and   \( \left\{ \left| \psi \right\rangle \otimes \left| \psi
\right\rangle, \left| \phi _{1}\right\rangle \otimes \left|  \phi
_{2}\right\rangle  \right\} \).  Since both direct individual LOCC
transformations,  and the collective LOCC transformation  are ruled out, we
require either two different catalyst states, one for each  pair, or a single
catalyst that can work for both the transformations.    Suppose the
entanglement supplier fails to provide  two catalysts for the two pairs or a
common catalyst that may work for both  of them, but instead provides only
one, say \( \left| \chi \right\rangle =\left( 0.625, 0.375\right)  \)  which
is useful only to carry out a single transformation, i.e.,  \begin{eqnarray} 
\left| \psi \right\rangle \otimes \left| \chi \right\rangle  & \rightarrow  &
\left| \phi _{1}\right\rangle \otimes \left| \chi  \right\rangle \ , \label{5}
\\  \left| \psi \right\rangle \otimes \left| \chi \right\rangle  &
\nrightarrow  & \left| \phi _{2}\right\rangle \otimes \left| \chi 
\right\rangle \ .\label{6}  \end{eqnarray}  It will be clear from the
following discussions as to why the given catalyst state doesn't work for  the
second transformation: It is not  entangled enough. In situations like this 
supercatalysis can provide a solution.    \emph{Step 1: (Supercatalysis)}
Perform a supercatalytic transformation involving the  incomparable pair
$\left\{ \ket{\psi},\ket{\phi_1}\right\}$ and the given auxiliary  state
$\ket{\chi}$:  \begin{equation} 
\label{12} 
\left| \psi \right\rangle \otimes \left| \chi \right\rangle \longrightarrow \left| \phi _{1}\right\rangle \otimes \left| \omega 
\right\rangle\ , 
\end{equation} 
where the new state is \( \left| \omega \right\rangle =\left( \frac{8}{13}, \frac{5}{13}\right)  \), 
with \( E\left( \omega \right) >E\left( \chi \right)  \). 
 
\emph{Step 2: (Catalysis)} The new improved auxiliary state, $\ket{\omega}$, is
now  sufficiently entangled to act as a legitimate catalyst for the second
incomparable  pair, and one can easily check that the transformation 
\begin{equation} 
\label{13} 
\left| \psi \right\rangle \otimes \left| \omega \right\rangle \longrightarrow \left| \phi _{2}\right\rangle \otimes \left|  
\omega 
\right\rangle 
\end{equation} 
can indeed be realized under LOCC with probability one. 
 
The above example shows that one might be able to perform a series of transformations 
with limited ancillary resources by improving the catalyst appropriately at 
every step to make it useful for subsequent transformations. 
 
In the rest of this letter,  we provide results on the existence of supercatalysts for given  
pairs of incomparable states, and study its relationship with catalyst states. For example,  
given a supercatalytic transformation what can we say about the ``catalytic'' properties of the auxiliary states?  
Clearly, if the two auxiliary states (i.e., the initial and the final auxiliary states)  
involved in the supercatalysis transformation 
are in \( 2\times 2 \),  then they are  {\em both} catalysts as well.  
However, whether such a property is always true for higher-dimensional auxiliary states 
is left as an open problem,  and the following result provides a sufficient condition.\vspace*{-1.5ex} 
\begin{pr}  Let $\ket{\chi}$ and $\ket{\phi}$ be 
the initial and final entangled states facilitating supercatalysis of the  
incomparable pair \( \left\{ \left| \psi \right\rangle ,\left| \phi \right\rangle \right\}  
\). If \( \left| \omega \right\rangle \rightarrow \left| \chi \right\rangle  \) 
under LOCC, then \( \left\{ \left| \chi \right\rangle ,\left| \omega \right\rangle \right\}  \) 
are also  catalysts for the incomparable pair \( \left\{ \left| \psi \right\rangle ,\left| \phi \right\rangle \right\}  \). 
\label{pr:supcat-to-cat}\vspace*{-1.5ex} 
\end{pr} 
{\bf Proof:} If \( \left| \omega \right\rangle \rightarrow \left| \chi \right\rangle  \), 
then we have the following transformations:  
(1) \( \left| \psi \right\rangle \otimes \left| \chi \right\rangle \rightarrow \left| \phi \right\rangle \otimes \left| \omega 
\right\rangle \rightarrow \left| \phi \right\rangle \otimes \left| \chi \right\rangle  \) 
and (2) \( \left| \psi \right\rangle \otimes \left| \omega \right\rangle \rightarrow \left| \psi \right\rangle \otimes \left| \chi 
\right\rangle \rightarrow \left| \phi \right\rangle \otimes \left| \omega \right\rangle  \) 
from which it follows that \emph{\( \left\{ \left| \chi \right\rangle ,\left| \omega \right\rangle \right\}  \)} 
are  catalysts for the incomparable pair \emph{}\( \left\{ \left| \psi \right\rangle ,\left| \phi \right\rangle \right\}  \). 
\qed  
 
As an immediate implication of the above proposition, we 
show the following bound on the entanglement of the final auxiliary state, $\ket{\omega}$. 
 \vspace*{-1.5ex} 
\begin{co} 
For a given incomparable pair \( \left\{ \left| \psi \right\rangle ,\left| \phi \right\rangle \right\}  
\) in $n \times n$, let \( k\times k \) states \( \left\{ \left| \chi \right\rangle ,\left| \omega \right\rangle \right\}  \) 
be the corresponding supercatalysts (i.e., \( \left| \psi \right\rangle \otimes \left| \chi \right\rangle \rightarrow \left| \phi 
\right\rangle \otimes \left| \omega \right\rangle  \) 
with probability one under LOCC,  and \( E\left( \left| \omega \right\rangle \right) >E\left( \left| \chi \right\rangle \right)  
\)). The improved state \( \left| \omega \right\rangle  \) can never be a maximally 
entangled state in \( k\times k \). \vspace*{-1.5ex} 
\end{co} 
{\bf Proof:} Let \emph{\( \left| \omega \right\rangle  \)} be a maximally entangled 
state in \( k\times k \). \emph{}Then  
\( \left| \omega \right\rangle \rightarrow \left| \chi \right\rangle  \). 
Therefore, by lemma 1, \( \left| \omega \right\rangle \textrm{ and }\left|
\chi \right\rangle  \)  are the catalysts for the given incomparable pair. But
a maximally entangled  state cannot be a catalyst \cite{jp}. Hence the proof.
\qed    
We next investigate the presence of supercatalysis when there exist catalytic states for 
a given pair of incomparable parent states. The associated  {\em formalism turns out to be 
extremely useful}: It provides a general framework and a necessary and sufficient condition 
for constructing  catalytic states, leads to 
sufficient conditions for supercatalysis and allows us to  
determine meaningful bounds on the enhanced entanglement of the auxiliart
state.   Given an incomparable pair  
\( \left\{ \left| \psi \right\rangle ,\left| \phi \right\rangle \right\}  \),  
with eigenvalue vectors \( \lambda _{\psi }=\left\{ \alpha _{1},\alpha _{2},\ldots 
,\alpha _{n}\right\} \textrm{ and }\lambda _{\phi }=\left\{ \beta _{1},\beta _{2},\ldots ,\beta _{n}\right\}  \), 
let \( \left| \chi \left( {\cal P}\right) \right\rangle  \) 
be a \( k\times k \) catalyst with the eigenvalue vector \( \displaystyle \lambda _{\chi }= {\cal P}= 
\left\{ p_{1},p_{2},\ldots ,p_{k}=1- \sum_{i=1}^{k-1} p_i \right\}  \), where 
$p_1\geq p_2\geq \ldots \geq p_{k-1}\geq p_k$. The proof of the following lemma provides a 
constructive computational procedure for determining all possible such $k\times  k$ catalytic states.  
\vspace*{-1.5ex} 
\begin{lm} 
The set of all $k\times k$ catalytic states for any given $n \times n$ pair of incomparable 
states,  \( \left\{ \left| \psi \right\rangle ,\left| \phi \right\rangle \right\}  \),  
is either empty, or a union of a finite number of  polyhedra in dimension $\leq (k-1)$.  
\label{lm:const-cat}\vspace*{-1.5ex} 
\end{lm} 
{\bf Proof:} Since we 
want auxiliary states, \(  \left| \chi \left( p_{1},p_{2},\ldots ,p_{k}=1- \sum_{i=1}^{k-1} p_i \right) \right\rangle  \),  
such that \( \lambda _{\psi \otimes \chi ({\cal P})}\prec \lambda _{\phi \otimes \chi ({\cal P})} \), the set of  
all possible ${\cal P}=\left(p_1, p_2, \ldots ,p_{k-1}\right)$ for which the auxiliary state is 
a catalytic state can be found as  
follows.  (i) Fix one possible ordering of the Schmidt coefficients of $\ket{\psi} \otimes \ket{\chi ({\cal P})}$, and determine 
the set of all possible ${\cal P}$ that satisfies this ordering by solving the  underlying  
$nk$ linear inequalities.  Hence, the set of ${\cal P}$ that correspond to a feasible 
fixed ordering of the eigenvalues of  
$\ket{\psi} \otimes \ket{\chi ({\cal P})}$, is a polyhedron (if an ordering is not feasible  for 
any choice of  ${\cal P}$, then the 
corresponding polyhedron is an empty set): the solutions 
of a set of linear inequalities defines a polyhedron. Also note that there are only a finite 
number of possible orderings of the eigenvectors of $\ket{\psi} \otimes \ket{\chi ({\cal P})}$, leading 
to a finite number of corresponding polyhedra: 
${\cal O}_1^\psi, {\cal O}_2^\psi, \ldots, {\cal O}_L^\psi$. An accurate 
estimate of $L$ can be obtained by viewing the counting problem as the number of possible ways $k$ sorted lists, 
each of length $n$, can be merged to generate distinct sorted lists of length $nk$; an upper bound on  
it is $(nk)!$. (ii) Similarly, compute the polyhedron for each  ordering of the eigenvalues of  
$\ket{\phi} \otimes \ket{\chi ({\cal P})}$. Again, this yields at most ${\cal O}_1^\phi, {\cal O}_2^\phi,  
\ldots, {\cal O}_L^\phi$ polyhedra.  
 
Now consider all possible polyhedra that are  
the intersections of pairs of non-empty order-preserving  
polyhedra defined above, i.e., $O_k= {\cal O}_i^\psi \cap {\cal O}_j^\phi$, 
$1\leq i,j\leq L$. The set of all points  in any such polyhedron $O_k$ that correspond to {\em catalytic states}, 
consists of  those points in $O_k$ that satisfy the underlying 
$nk-1$ majorization  linear inequalities (see Eq.~\ref{eq:majorization}):  $\lambda_{\psi \otimes \chi ({\cal P})}  
\prec \lambda_{\psi \otimes \chi ({\cal P})}$. Hence, the catalytic states within $O_k$ 
forms a polyhedron itself. Thus, each  polyhedron representing  
values of ${\cal P}$ that correspond to  catalytic states for the given 
pair \( \left\{ \left| \psi \right\rangle ,\left| \phi \right\rangle \right\}  \), can be viewed as 
 the intersections of three different polyhedra: (i) the set of ${\cal P}$ corresponding to a  
 fixed ordering of the Schmidt coefficients of $\ket{\psi} \otimes \ket{\chi ({\cal P})}$,  
 (ii) the set of ${\cal P}$ corresponding to a  fixed ordering of the  
 Schmidt coefficients of $\ket{\phi} \otimes \ket{\chi ({\cal P})}$, 
 and (iii) the set of all ${\cal P}$  
 that satisfy the majorization relations corresponding to the fixed 
 orderings defined in (i) and (ii).   
 We define such a polyhedron (which is the intersection of the 
 preceding three polyhedra) as an {\em Order Preserving Majorization Polyhedron}  (OPMP).  
\qed 
 
For catalytic states in any dimension $k\times k$, a typical OPMP, $S_i$, can be represented by the 
extreme points (or vertices) of the underlying polyhedron: $S_i=\left\{ {\cal P}_1, {\cal P}_2, 
\ldots , {\cal P}_m\right\}$, where ${\cal P}_i \in {\cal R}^{k-1}$, and 
$E(\ket{\chi({\cal P}_1)}) \geq E(\ket{\chi({\cal P}_2)})\geq \ldots 
\geq  E(\ket{\chi({\cal P}_m)})$. For example, for $k=2$, one can represent each OPMP as an interval 
belonging to the segment $\left[ \frac{1}{2}, 1\right]$:  
\( S=\left[ p_{l},p_{u}\right]  \), where \( E\left( \left| \chi (p_{l})\right\rangle \right) >E\left( \left| \chi 
(p_{u})\right\rangle \right)  \). By following the procedure outlined in the proof of the preceding lemma, 
it is fairly easy to construct all OPMPs for any given catalyzable incomparable 
pair, especially for small values of $n$ and $k$. For instance, an 
OPMP for the states given by Eqs. (\ref{eq:1}) and (\ref{eq:2}) is: \( S_{1}=\left[ 
\frac{10}{19},\frac{25}{38}\right]  \). 
Another OPMP for the same pair but corresponding to a different ordering is: \( S_{2}=\left[ 
\frac{13}{25},\frac{10}{19}\right]  \).  
 
The framework introduced in Lemma~\ref{lm:const-cat} shows for the first time that the 
set of all possible catalysts can be structured in terms of a {\em discrete and a finite} 
number of polyhedra, each of which has an efficient description (i.e., the corresponding  
vertices). Hence, our  framework   
provides a {\em succinct necessary and sufficient condition} for determining whether 
a given pair of incomparable states is catalyzable or not, as captured in the following theorem.  
\vspace*{-4ex} 
\begin{theo} 
A given $n\times n$ incomparable pair of states is catalyzable if and only if there exists a 
non-empty OPMP in some $k\times k$. 
\label{theo:cat-nec-suff}\vspace*{-1.5ex} 
\end{theo} 
Note that the computational problem for finding catalysts  
(i.e., given a pair of incomparable states in 
$n \times n$, does there exist a catalytic state  in $k \times k$? ) is in the
class NP \cite{NP}: in order  to provide a valid certificate for a ``yes"
instance of the problem, all one needs to do  is to provide a candidate
catalytic state, $\ket{\chi}$, and one can verify in $O(nk)$ time   whether
$\ket{\chi}$ is indeed a catalytic state or not. Lemma~\ref{lm:const-cat} and  
Theorem~\ref{theo:cat-nec-suff} provide an $O([(nk)!]^2)$ algorithm not only
to solve the  ``yes/no" version of the problem, but also to determine all the
possible catalytic states.   Whether the catalysis problem admits an efficient
solution, or is an NP-complete problem,   is left as an open problem.    
The preceding understanding of the structure of 
catalytic states can now be used to establish a connection between catalysis and supercatalysis 
and establish a sufficient condition for the latter. First we introduce certain structures 
of the majorization relations. A {\em parameterized} majorization relationship,   
$\lambda _{\psi \otimes \chi ({\cal P})}\prec \lambda _{\phi \otimes \chi ({\cal P})} \), where 
${\cal P}=\left(p_1, p_2, \ldots ,p_{k-1}\right)$, is said to be {\em strict} if there 
exists an OPMP of dimension $\geq 1$ (i.e., it is non-empty and is not a single point), 
such that  there exists a point ${\cal P}_1$ in the OPMP where {\em all} the nontrivial $(nk)-1$,  
 majorization inequalities  (see Eq.~\ref{eq:majorization}) are strict.  We represent strict majorization as  
 $\lambda _{\psi \otimes \chi ({\cal P})} \sqsubset \lambda _{\phi \otimes \chi ({\cal P})}$. Moreover,   
 a parameterized majorization relationship, $\lambda _{\psi \otimes \chi ({\cal P})}\prec \lambda _{\phi \otimes \chi ({\cal P})} \), is said to be {\em semi-strict} if there 
exists an OPMP of dimension $\geq 1$ (i.e., it includes at least a line segment), 
such that  there exist a point ${\cal P}_1$ in the OPMP and a direction vector $\vec{d} \in {\cal R}^{k-1}$ 
such that ${\cal P}_1-\epsilon \vec{d}$ is also in the OPMP, and  
any  equality relations in the majorization relationship at ${\cal P}_1$ holds even if ${\cal P}_1$ is 
replaced by ${\cal P}_1-\epsilon \vec{d}$ on the {\em right-hand side}; we refer to such equalities 
in the majorization relationships as {\em benign} \cite{som}. Note that {\em strict majorization 
is a special case of the semi-strict case}, and we  represent semi-strict majorization as  
 $\lambda _{\psi \otimes \chi ({\cal P})} \sqsubseteq \lambda _{\phi \otimes \chi ({\cal P})}$  
 \cite{som}. Note also that since $E(\ket{\chi({\cal P})})$ is a concave function, then without 
 loss of generality,  we can assume that it {\em increases} along the direction $-\vec{d} \in {\cal R}^{k-1}$ (if 
 not, then just reverse the sign of $\vec{d}$). \vspace*{-1.5ex} 
\begin{theo} 
Given an $n\times n$ catalyzable incomparable pair  
$ \left\{ \left| \psi \right\rangle ,\left| \phi \right\rangle \right\}$ that admits catalysts in $k \times k$, 
{\em supercatalysis also occurs in $k\times k$} for the given incomparable pair 
if   $\lambda _{\psi \otimes \chi ({\cal P})} \sqsubseteq \lambda _{\phi \otimes \chi ({\cal P})}$.  
\label{theo:cat-to-supcat}\vspace*{-1.5ex} 
\end{theo} 
{\bf Proof:}  Since  
$\lambda _{\psi \otimes \chi ({\cal P})} \sqsubseteq \lambda _{\phi \otimes \chi ({\cal P})}$,  
then it follows from the preceding definitions  that there exist ${\cal P}_1, \vec{d} \in {\cal R}^{k-1}$ and 
an $\epsilon >0$ such that $\lambda _{\psi \otimes \chi ({\cal P_1})} \prec 
 \lambda _{\phi \otimes \chi ({\cal P}_1 -\epsilon \vec{d})}$. The proof is direct: 
 first pick a valid direction vector $\vec{d}$ and an $\epsilon$ small enough so that   
 ${\cal P}_2= {\cal P}_1 -\epsilon \vec{d}$ is 
 still in the OPMP and all the majorization inequalities are still 
 satisfied when ${\cal P}_2$ is used for the right-hand side of the majorization 
 inequalities. Moreover, note that the entropy function increases along the direction $-\vec{d}$. Hence,  
 to obtain supercatalysis, set $\ket{\chi}= \ket{\chi({\cal P}_1)}$ as the initial entangled state 
 and $\ket{\omega}=\ket{\chi ({\cal P}_1 -\epsilon \vec{d})}$ as the final 
 auxiliary entangled state. \qed 
  
 We next discuss the amount by which  the entanglement of the auxiliary 
state can be enhanced by using the constructive procedure stated in Theorem~\ref{theo:cat-to-supcat}.   
In other words, we would like to maximize the enhancement 
\( \delta =E\left( \omega \right) -E\left( \chi \right)  \), because by doing so  
the overall loss of entanglement in the transformation is minimized. 
 In the procedure of Theorem~\ref{theo:cat-to-supcat}, since both $\ket{\chi}$ and $\ket{\omega}$ 
belong to the same OPMP, say $S=\left\{ {\cal P}_1, {\cal P}_2, 
\ldots , {\cal P}_m\right\}$ (recall that the vertices of the OPMP 
are ordered in terms of decreasing entanglement), then the maximum enhancement,  
$\delta \leq E\left( \left| \chi ({\cal P}_{1})\right\rangle \right) -E\left( \left| \chi ({\cal P}_{m})\right\rangle 
\right) $.  Take for instance, one of the OPMP's for the states in  
Eqs. (\ref{eq:1}) and (\ref{eq:2}): \( S_{1}=\left[ 
\frac{10}{19},\frac{25}{38}\right]  \). If we choose \( \left| \omega \right\rangle =\left| \chi \left( \frac{10}{19}\right) \right\rangle  \) 
and \( \left| \chi \right\rangle =\left| \chi \left( \frac{25}{38}\right) \right\rangle  \) 
then one can check that the transformation \( \left| \psi \right\rangle \otimes \left| \chi \right\rangle \nrightarrow \left| \phi 
\right\rangle \otimes \left| \omega \right\rangle  \) 
is not possible with certainty by LOCC.  This shows that the preceding upper bound  
on the enhanced entanglement is not always attained. However, one can verify that the  
conditions of Theorem~\ref{theo:cat-to-supcat} are satisfied by $S_1$, and that one   
find two catalyst states in $S_1$ such that supercatalysis does indeed happen. Next, 
consider another OPMP for the same incomparable pair: \( S_{2}=\left[ 
\frac{13}{25},\frac{10}{19}\right]  \).  In this case, one can easily prove that the upper bound 
is indeed attained. It is clear that the amount of enhancement depends on the choice 
of OPMP. An optimal strategy would be to consider all possible OPMPs and to obtain 
the optimal pair that belongs to one particular OPMP for supercatalysis. This 
is however beyond the scope of this letter. 
 
We now come to the question of {\em efficiency of supercatalysis}. The dimension of 
the auxiliary state, $\ket{\chi}$, plays a crucial role in determining the complexity and efficiency 
of an entanglement assisted transformation. To reduce complexity and increase 
efficiency, it is necessary to keep the dimension of the borrowed entanglement 
at a minimum whenever possible. Theorem~\ref{theo:cat-to-supcat} provides 
sufficient conditions where catalysis leads to supercatalysis, 
{\em without increasing the dimension} of the auxiliary entangled states.  
However, we show next that  there exist cases where catalysts exist in $k\times k$,   
but  supercatalysis can never happen without increasing the dimension of the 
auxiliary states. Consider  
the following incomparable parent states in $5 \times 5$: $\psi= 
\left(.4,.3,.2,.05,.05 \right)$, and $\phi=\left(.4,.35,.14,.11,0\right)$. 
One can verify that this incomparable pair admits a catalyst, $\ket{\chi}=\left(0.6,0.4\right)$. 
The following 
theorem, however, shows that the parent incomparable states {\em cannot participate} in any  
supercatalysis, without increasing the dimension of the entangled states to $\geq 3$.  
\vspace*{-1.5ex} 
\begin{theo} 
Let \( \left\{ \left| \psi \right\rangle ,\left| \phi \right\rangle \right\}  \) 
be an incomparable pair with eigenvalue vectors  \( \lambda _{\psi }=\left\{ \alpha _{1},\alpha _{2},\ldots ,\alpha 
_{n}\right\} ,\lambda _{\phi }=\left\{ \beta _{1},\beta _{2},\ldots ,\beta _{n}\right\}  \). 
If \( \alpha _{1}=\beta _{1} \) or \( \alpha _{n}=\beta _{n} \) then 
{\em supercatalysis is not possible with \( 2\times 2 \) auxiliary states}. Moreover 
if \( \alpha _{1}=\beta _{1} \) and \( \alpha _{n}=\beta _{n} \) then there 
are no \( 3\times 3 \) auxiliary states for supercatalysis. \vspace*{-1.5ex} 
\end{theo} 
\textbf{Proof:} Let there exist an auxiliary entangled state \( \left| \chi \right\rangle  \) 
such that \( \left| \psi \right\rangle \otimes \left| \chi \right\rangle \rightarrow \left| \phi \right\rangle \otimes \left| \omega 
\right\rangle  \) 
where \( E\left( \left| \omega \right\rangle \right) >E\left( \left| \chi \right\rangle \right)  \)\emph{.} 
Let \( \lambda _{\chi }=\left\{ p,1-p\right\} ,\lambda _{\omega }=\left\{ q,1-q \right\}  \). 
Since \emph{\( E\left( \left| \omega \right\rangle \right) >E\left( \left| \chi \right\rangle \right)  \)} 
therefore \emph{\( p>q \)}. Since \( \left| \psi \right\rangle \otimes \left| \chi \right\rangle \rightarrow \left| \phi 
\right\rangle \otimes \left| \omega \right\rangle  \), 
we have \( \alpha _{1}p\leq \beta _{1}q \). Since \( \alpha _{1}=\beta _{1} \) 
therefore, \( p\leq q \) which is a contradiction. Similar proof for the case 
when \( \alpha _{d}=\beta _{d} \). 
 
To prove the second part of the lemma assume there are \emph{\( 3\times 3 \)} 
auxiliary states \( \left| \chi \right\rangle  \) and \( \left| \omega \right\rangle  \) 
such that \( \left| \psi \right\rangle \otimes \left| \chi \right\rangle \rightarrow \left| \phi \right\rangle \otimes \left| \omega 
\right\rangle  \) 
where \( \left| \chi \right\rangle \rightarrow \left| \phi \right\rangle \otimes \left| \omega \right\rangle  \). 
Let \( \lambda _{\chi }=\left\{ p_{1},p_{2},1-p_{1}-p_{2}\right\} \textrm{ and }\lambda _{\omega }=\left\{ 
q_{1},q_{2},1-q_{1}-q_{2}\right\}  \). 
We then have \( \alpha _{1}p_{1}\leq \beta _{1}q_{1}\Rightarrow p_{1}\leq q_{1}\textrm{since }\alpha _{1}=\beta 
_{1} \) 
and \( 1-\alpha _{n}\left( 1-p_{1}-p_{2}\right) \leq 1-\beta _{n}\left( 1-q_{1}-q_{2}\right) \Rightarrow 
p_{1}+p_{2}\leq q_{1}+q_{2} \). 
Hence, $\lambda_\chi \prec \lambda_\omega$ and  \( \left| \chi \right\rangle \rightarrow \left| \omega \right\rangle \Rightarrow E\left( \left| \omega \right\rangle \right) 
<E\left( \left| \chi \right\rangle \right)  \) (see \cite{nielsen})  
which is a contradiction.  \qed 
 
What happens if one cannot obtain auxiliary states for supercatalysis in the same dimension 
as the catalysts?  Since the augmented pair $\{\ket{\psi}\otimes \ket{\chi}, 
\ket{\phi}\otimes \ket{\chi}$ is LOCC transformable, one can state the 
following result based on the results on recovery of entanglement in  
\cite{som}.  \vspace*{-1.5ex}  
\begin{theo} 
Let $\ket{\psi}=\left( \alpha_1, \alpha_2, \ldots, \alpha_{n}\right)$ and 
$\ket{\phi}=\left( \beta_1, \beta_2, \ldots, \beta_{n1}\right)$, be an
incomparable pair, where $\alpha_{n} \neq \beta_{n}$, and let the pair
admit a $k \times k$ catalyst, $\ket{\chi}$. Then the pair $\{ \ket{\psi}, 
\ket{\phi}\}$ admits  supercatalysts, with initial auxiliary state,
$\ket{\chi^\prime}= \ket{\chi} \otimes \ket{\chi_1}$, and  the final enhanced
auxiliary state, $\ket{\omega^\prime}= \ket{\chi} \otimes \ket{\omega_1}$,
where  $\ket{\chi_1}$ and $\ket{\omega_1}$ are in dimension $m\times m$, $m\leq
 nk-1$ and $E(\ket{\omega_1}) > E(\ket{\chi_1})$.  
\label{theo:cat-to-scat2}\vspace*{-1.5ex}   \end{theo}  To summarize, we have
shown the existence of entanglement assisted transformations  that are more
efficient than catalysis. In such transformations, called supercatalysis,  the
entanglement of the auxiliary state is enhanced at the end and therefore the
net loss  in entanglement is reduced. We obtained a set of sufficient
conditions for supercatalysis  to exist and explored several relationships
between supercatalysis and catalysis.  There are many open questions of
interest, including: What are some of   the necessary conditions for
supercatalysis? Are the auxiliary states participating in a supercatalysis
process also catalysts for the parent incomparable states? Is the existence of
catalysis always sufficient to ensure supercatalysis?  Are  the problems of
finding catalysts and supercatalysts for a given incomparable pair
NP-Complete?

The work  was sponsored in part by the Defense Advanced Research
Projects Agency (DARPA) project MDA 972-99-1-0017 (note that the content of
this paper does not necessarily reflect the position or the policy of the government
and no official endorsement should be inferred), and in part by the U.S. Army
Research Office/DARPA under contract/grant number DAAD 19-00-1-0172.

 \end{document}